\documentclass[prd,showpacs,aps,preprintnumbers]{revtex4}
\usepackage[mathscr]{eucal}
\usepackage{color}
\usepackage{amsmath}
\usepackage{graphicx}
\usepackage{bm}

\newcommand{\be}{\begin{equation}}

\newcommand{\ee}{\end{equation}}
\newcommand{\bea}{\begin{eqnarray}}
\newcommand{\eea}{\end{eqnarray}}

\def\slr#1{\setbox0=\hbox{$#1$}           % set a box for #1
   \dimen0=\wd0                                 % and get its size
   \setbox1=\hbox{/} \dimen1=\wd1               % get size of /
   \ifdim\dimen0>\dimen1                        % #1 is bigger
      \rlap{\hbox to \dimen0{\hfil/\hfil}}      % so center / in box
      #1                                        % and print #1
   \else                                        % / is bigger
      \rlap{\hbox to \dimen1{\hfil$#1$\hfil}}   % so center #1
      /                                         % and print /
   \fi}

\def\be{\begin{eqnarray}}
\def\ee{\end{eqnarray}}

\begin{document}
\preprint{BARI-TH 536/06}

\author{R.Casalbuoni$^{a,b}$}
\author{M.Ciminale$^{d,e}$}
\author{R.Gatto$^c$}
\author{G.Nardulli$^{d,e}$}
\author{M.Ruggieri$^{d,e}$}

\affiliation{$^a$Dipartimento di Fisica, Universit\`a di Firenze,
I-50125 Firenze, Italia} \affiliation{$^b$I.N.F.N., Sezione di
Firenze, I-50125 Firenze, Italia} \affiliation{$^c$D\'epart. de
Physique Th\'eorique, Universit\'e de Gen\`eve, CH-1211 Gen\`eve 4,
Suisse} \affiliation{$^d$Dipartimento di Fisica, Universit\`a di
Bari, I-70126 Bari, Italia}\affiliation{$^e$I.N.F.N., Sezione di
Bari, I-70126 Bari, Italia}
\date{\today}

\title{Influence of finite quark chemical potentials on the three flavor LOFF
phase of QCD}
\date{\today}

\begin{abstract}
We study in the Ginzburg-Landau approximation,  the
Larkin-Ovchinnikov-Fulde-Ferrell (LOFF) phase of QCD with three
flavors and one plane wave, including terms of order ${\cal
O}(1/\mu)$. We show that the LOFF window is slightly enlarged, and
actually splits into two different regions, one characterized by
$u-s$ and $d-u$ pairings and the other with $d-u$ pairs only.
\end{abstract}

%\pacs{12.38.Aw, 12.38.Lg}


\maketitle

\section{Introduction}

At high densities and small temperatures quarks are expected to
attract each other in the color antisymmetric channel. Cooper pairs
are expected to be formed and color superconductivity to arise,
see~\cite{barrois,Alford:1997zt} and \cite{Rajagopal:2000wf} for
reviews of the subject. For three quark flavors, the
Color-Flavor-Locking (CFL) phase is known to be the ground state,
provided the density is asymptotically high. The condensate is a
spinless color- and flavor- antisymmetric
diquark~\cite{Alford:1998mk}.

Pre-asymptotic densities and $T\simeq 0$ are probably  relevant to
describe the cores of compact stars (we will consider only the zero
temperature case in this paper). At those densities one cannot
neglect the mass of the strange quark and the chemical potential
differences $\delta\mu$ due to $\beta$ equilibrium. Much effort has
been devoted to study phases which might be relevant at such
pre-asymptotic densities, such as the 2SC
phase~\cite{Alford:1997zt}, the gapless phases
g2SC~\cite{Shovkovy:2003uu}, and, most remarkably, the gapless gCFL
phase \cite{Alford:2003fq,Alford:2004hz}. The gapless gCFL phase was
considered for some time as the most suitable candidate. It was
however realized that imaginary gluon Meissner masses (for g2SC
see~\cite{Huang:2004bg}, and in particular for the important
candidate gCFL see~\cite{Casalbuoni:2004tb}) originate instabilities
in the gapless phases (and also the 2SC phase shows
instability~\cite{Huang:2004bg,Huang:2005pv,Gorbar:2005rx}). The
possibility to remove the instability from the g2SC phase  has been
recently discussed in Ref. \cite{Giannakis:2006gg}.

 At all events, for non-zero differences of
the chemical potentials, quark pairings with total non-vanishing
momenta may take place, see~\cite{Alford:2000ze} and for a
review~\cite{Casalbuoni:2003wh}. These pairings are called LOFF
pairings, from  old superconductivity studies by
Larkin-Ovchinnikov-Fulde-Ferrell (LOFF)~\cite{LOFF2}. LOFF pairing
gives rise to LOFF phases. In the case of two flavors it has been
shown that instability in 2SC indicates that the LOFF phase is
energetically favored ~\cite{Giannakis:2004pf}, provided there is
chromomagnetic stability in the LOFF phase
~\cite{Giannakis:2005vw,Gorbar:2005tx}.

The interesting LOFF case for physics is however that of three
flavors. In preliminary studies for three flavors it was found, in
the Ginzburg-Landau (GL) approximation, that condensation of the
$u-s$ and $d-u$ pairs is possible in the form of inhomogeneous LOFF
pairing \cite{Casalbuoni:2005zp}, and subsequently it was found that
such a phase is chromomagnetic stable \cite{Ciminale:2006sm}. The
validity of the G-L approximation for such a crystal structure has
recently been tested \cite{Mannarelli:2006fy}. The particular LOFF
phase studied in these papers suggested therefore that LOFF phases,
in general, could remedy at the "impasse" originated from the
discovered chromomagnetic instability of gCFL. These studies where
made for the leading terms in the inverse of the baryon chemical
potential and the problem remained to check the validity of such
approximation. At the same time the color chemical potentials had
been neglected on the basis of previous experience on the subject,
but without quantitative check. In the present paper we complete
such investigations by going beyond these approximations.

Recently an important study for three flavors has been completed by
Rajagopal and Sharma \cite{Rajagopal:2006ig,Rajagopal:2006dp},
always within the G-L approximation, but including the study of
structures higher than the simplest crystal structure studied by us.
The authors come out with two preferred structures with
face-centered cubic symmetry, one with separate cubic structure of
$u-s$ and $d-u$, the other with combined cubic structure for both.
Besides the G-L approximation, they make use, in their very complex
study, of some approximations that we try to overcome in the present
paper. Although we consider a simpler crystal structure, our results
could be relevant also for more complex crystalline structures. The
other independent aspect, i.e. the validity of the G-L
approximation, still remains to be solved, and it will probably
require long efforts to definitely clarify the nature of the stable
phases of QCD at intermediate densities under neutrality conditions
and non negligible strange quark mass. But the importance of
comparing with the LOFF phases seems unavoidable at this stage of
the knowledge, and LOFF phases remain as very serious candidates for
the solution of QCD at non-asymptotic densities.

The plan of the paper is as follows: In section \ref{sec2} we review
the three flavor LOFF phase of QCD. In section \ref{sec3} we make
the $1/\mu$ expansion. In section  \ref{sec4} we discuss our
results. Section \ref{sec5} contains our conclusions.

\section{Review of the three flavor LOFF phase of QCD\label{sec2}}
In this section we  briefly review the main elements for the study
of the LOFF phase. The Lagrangean density for three flavor ungapped
quarks is:
\begin{equation}
{\cal L}=\bar{\psi}_{i\alpha}\,\left(i\,D\!\!\!\!
/^{\,\,\alpha\beta}_{\,\,ij} -M_{ij}^{\alpha\beta}+
\mu^{\alpha\beta}_{ij} \,\gamma_0\right)\,\psi_{\beta j}
\label{lagr1}\ .
\end{equation}
 where $M_{ij}^{\alpha\beta} =\delta^{\alpha\beta}\, {\rm diag}(0,0,M_s)
$ is the mass matrix and
$D^{\alpha\beta}_{ij}=\partial\delta^{\alpha\beta}\delta_{ij}+
igA_aT_a^{\alpha\beta}\delta_{ij}$. The indices $i,\,j$ refer to
flavor and $\alpha$, $\beta$ to color. The matrix
$\mu_{\alpha\beta}^{ij}$ is a diagonal matrix, which  depends on
$\mu$ (the average quark chemical potential), $\mu_e$ (the electron
chemical potential), and $\mu_3,\,\mu_8$ (color chemical potentials)
\cite{Alford:2003fq}. It can be written as follows\be
{\bf\mu}_{ij}^{\alpha\beta} = \left(\mu \delta_{ij}- \mu_Q
Q_{ij}\right)\delta^{\alpha\beta}+ \delta_{ij} \left(\mu_3
T_3^{\alpha\beta}+\frac{2}{\sqrt 3}\mu_8 T_8^{\alpha\beta}\right)
\label{9}\ee ($i,j =1,3 $ flavor indices; $\alpha,\beta =1,3 $ color
indices). Here $T_3 = \frac 1 2 {\rm diag}(1,-1,0)$, $T_8 =
\frac{1}{2 \sqrt 3 }{\rm diag}(1,1,-2)$ in color space and $Q= {\rm
diag} (2/3,-1/3,-1/3)$ in flavor space; $ \mu_e$ is the electron
chemical potential; $\mu_3, \mu _8$ are the color chemical
potentials associated respectively to the color charges $T_3$ and
$T_8$; $\mu$ is the baryonic chemical potential which we fix to
$500$ MeV. One has the following chemical potentials for the nine
different quarks:\bea
\mu_{u_r}&=&\mu-\frac{2\mu_e}{3}+\frac{\mu_3}{2}+\frac{\mu_8}{3}
~,~~~~~~~
\mu_{u_g}=\mu-\frac{2\mu_e}{3}-\frac{\mu_3}{2}+\frac{\mu_8}{3}~,~~~
~~~~ \mu_{u_b}=\mu-\frac{2\mu_e}{3}-\frac{2\mu_8}{3}~, \cr
\mu_{d_r}&=&\mu+\frac{\mu_e}{3}+\frac{\mu_3}{2}+\frac{\mu_8}{3}
~,~~~~~~~
\mu_{d_g}=\mu+\frac{\mu_e}{3}-\frac{\mu_3}{2}+\frac{\mu_8}{3}~,~~~
~~~~ \mu_{d_b}=\mu+\frac{\mu_e}{3}-\frac{2\mu_8}{3}~,\cr
\mu_{s_r}&=&\mu+\frac{\mu_e}{3}+\frac{\mu_3}{2}+\frac{\mu_8}{3}~,~~~~~~~
\mu_{s_g}=\mu+\frac{\mu_e}{3}-\frac{\mu_3}{2}+\frac{\mu_8}{3}~,~~~~~~~
\mu_{s_b}=\mu+\frac{\mu_e}{3}-\frac{2\mu_8}{3}\ .
 \label{11}\eea

We make use of the  High Density Effective Theory (HDET), see
\cite{Hong:1998tn,Beane:2000ms,Casalbuoni:2003cs} and for reviews
 \cite{Rajagopal:2000wf}. The quark momentum is
written as the sum of a large component $\mu{\bf n}$, where ${\bf n
}$ is a unit vector and a residual small component $\ell$. We also
introduce $\bf n$-dependent fields $\psi_{\bf n} $ and $\Psi_{\bf
n}$ by
\begin{equation}
\psi(x)=\int\frac{d{\bf n}}{4\pi}e^{i\,\mu{\bf n}\cdot{\bf
x}}\,\left(\psi_{\bf n}(x)+\Psi_{\bf n}(x)\right)\label{decomp} \,;
\end{equation}
here $\psi_{\bf n}$ and $\Psi_{\bf n}$  correspond respectively to
positive and negative energy Dirac solutions.

We change the spinor basis by defining $\psi_A = (\psi_{ur},
 \psi_{dg},\psi_{bs},\psi_{dr},\psi_{ug},
 \psi_{sr},\psi_{ub},\psi_{sg},\psi_{db})$.
This is done through unitary matrices $F_A$, which are reported in
Ref. \cite{Casalbuoni:2004tb}. Finally a Nambu-Jona Lasinio four
fermion coupling is added to the Lagrangean. This term is taken in
the mean field approximation. The procedure corresponds to the same
coupling and the same approximation as in Ref. \cite{Alford:2004hz}.
Next, one introduces the Nambu-Gorkov field
\begin{equation}
\chi_A=\frac{1}{\sqrt{2}}\left(\begin{array}{c}
  \psi_{\bf n} \\
  C\,\psi^*_{- \bf n}
\end{array}\right)_A \label{nambu-gorkov}\ ,
\end{equation} and the Lagrangean can be written in the form
\begin{equation}
L=\frac{1}{2}\sum_{A,B}\int\frac{d{\bf
n}}{4\pi}\,\int\frac{dE\,d\xi}{(2\pi)^2}\,\chi^\dagger_A\,\left(
\begin{array}{cc}
  \left(E-\xi + \bar\mu_{A}\right)\delta_{AB} & -\Delta_{AB}({\bf r}) \\
  -\Delta_{AB}^*({\bf r}) &  \left(E+\xi - \bar\mu_{A} \right)\delta_{AB}
\end{array}
\right)\,\chi_B \label{kinetic}
\end{equation}
with $ (\bar\mu)_{A} \ = \ \left(\bar\mu_{u}, \bar\mu_{d},
\bar\mu_{s}, \bar\mu_{d}, \bar\mu_{u}, \bar\mu_{s}, \bar\mu_{u},
\bar\mu_{s}, \bar\mu_{d}\right)$. In the above equation $E$ is the
energy, $\xi\equiv {\bf \ell}\cdot{\bf n}$ is the component of the
residual momentum along $\bf n$ which satisfies $|\xi|<\delta$, with
$\delta$ an ultraviolet cutoff.

We make a Fulde-Ferrell ansatz for each inhomogeneous pairing. The
ansatz is
\begin{equation}
<\psi_{i\alpha}\,C\,\gamma_5\,\psi_{\beta j}> =
\sum_{I=1}^{3}\,\Delta_I({\bf r})\,\epsilon^{\alpha\beta
I}\,\epsilon_{ijI}~\label{cond}
\end{equation}
with \be \Delta_I ({\bf r}) = \Delta_I \exp\left(2i{\bf
q_I}\cdot{\bf r}\right)~, \label{eq:1Ws}\ee where $2{\bf q_I}$ is
the the Cooper pair momentum. In the gap matrix $\Delta_{AB}$ in
(\ref{kinetic}) there are three independent gap functions
$\Delta_{1}({\bf r})$, $\Delta_{2}({\bf r})$, $\Delta_{3}({\bf r})$.
They correspond respectively to $d-s$, $u-s$ and $u-d$ pairing. The
matrix $\Delta_{AB}$ is given explicitly in \cite{Alford:2003fq},
\cite{Casalbuoni:2004tb}.

To determine the vectors ${\bf q_I}$ one has to minimize the free
energy. The norms $|{\bf q_I}|$ can be determined by the
minimization procedure. As for the directions, one separately
analyzes different structures to find out which is the favorite one.
In our previous paper \cite{Casalbuoni:2005zp} we had neglected the
color potentials since they vanish in the normal phase and neglected
the \cal{O}(${1}/{\mu}$) corrections. The favored structure was
found to be that with the vectors ${\bf q_2}$ and ${\bf q_3}$
parallel. In the subsequent work \cite{Ciminale:2006sm} it was found
that such a phase is chromomagnetic stable. A step forward came from
the work of Mannarelli, Rajagopal and Sharma
\cite{Mannarelli:2006fy} who went beyond the G-L approximation
confirming its approximate validity and the parallel situation of
the two vectors in the favorite phase. Here we shall include the
\cal{O}(${1}/{\mu}$) corrections and study the role of the color
chemical potentials.

\section{$1/\mu$ expansion\label{sec3} }

We assume from the very beginning $\Delta_1 = 0$ and ${\bf\hat q}_2
= {\bf\hat q}_3$ (the hat denotes a unit vector). However we here
include the chemical potentials $\mu_3$ and $\mu_8$. We consider
therefore the Ginzburg-Landau functional
\begin{equation}
\Omega = \Omega_0 + \frac{\alpha_2}{2}\Delta_2^2 +
\frac{\beta_2}{4}\Delta_2^4 + \frac{\alpha_3}{2}\Delta_3^2 +
\frac{\beta_3}{4}\Delta_3^4 + \frac{\beta_{23}}{2}\Delta_2^2
\Delta_3^2  +\emph{O }(\Delta^6)~. \label{eq:Free}
\end{equation}
where the coefficients $\alpha_j$, $\beta_j$ and $\beta_{23}$ are
defined in the following way: \bea && \alpha_2 =
\alpha(q_2,\delta\mu_{u_r,s_b},\delta\mu_{u_b,s_r})~,~~~~~~~~~
\alpha_3 = \alpha(q_3,\delta\mu_{u_r,d_g},\delta\mu_{u_g,d_r})~, \\
&&\beta_2 =
\beta(q_2,\delta\mu_{u_r,s_b},\delta\mu_{u_b,s_r})~,~~~~~~~~~
\beta_3 = \beta(q_3,\delta\mu_{u_r,d_g},\delta\mu_{u_g,d_r})~, \\
&&\beta_{23}(q_2,q_3,\delta\mu_{ud},\delta\mu_{us}) =
-\frac{4\mu^2}{\pi^2}\frac{1}{4(q_2\cdot\delta\mu_{ud}+q_3\cdot\delta\mu_{us})}
\left(\log\left|\frac{q_3-\delta\mu_{ud}}{q_3+\delta\mu_{ud}}\right|-\log\left|\frac{q_2+\delta\mu_{us}}{q_2-\delta\mu_{us}}\right|\right)\eea
and \bea &&\alpha(q,\delta\mu_1,\delta\mu_2) =
\frac{1}{2}\alpha(q,\delta\mu_1) +
\frac{1}{2}\alpha(q,\delta\mu_2)~,~~\label{alphaTOT} \\ &&
\alpha(q,\delta\mu) = -\frac{4\mu^2}{\pi^2}\left(1 -
\frac{\delta\mu}{2q}\log\left|\frac{q+\delta\mu}{q-\delta\mu}\right|
- \frac{1}{2}\log\left|\frac{4(q^2
-\delta\mu^2)}{\Delta_0^2}\right|\right)~, \\
&&\beta(q,\delta\mu_1,\delta\mu_2) = \frac{1}{2}\beta(q,\delta\mu_1)
+ \frac{1}{2}\beta(q,\delta\mu_2)~,\label{betaTOT} \\ &&
\beta(q,\delta\mu) = \frac{\mu^2}{\pi^2}\frac{1}{q^2-\delta\mu^2}~.
\eea $\beta_{23}$ is related to the $3\times3$ block of the gap
matrix, therefore it depends only on
$\delta\mu_{ud}=\delta\mu_{u_r,d_g}$ and
$\delta\mu_{us}=\delta\mu_{u_r,s_b}$; the  chemical potentials
differences are given by\bea&& \delta\mu_{u_r,d_g} =
\frac{\mu_{d_g}-\mu_{u_r}}{2}~,~~~~
\delta\mu_{u_g,d_r} = \frac{\mu_{d_r}-\mu_{u_g}}{2}~,\nonumber\\
&&\delta\mu_{u_r,s_b} =
\frac{\mu_{u_r}-\mu_{s_b}}{2}~,~~~~\delta\mu_{u_b,s_r} =
\frac{\mu_{u_b}-\mu_{s_r}}{2}~. \label{eq:DefDiff}  \eea We expand
the free energy in the chemical potentials ($ \mu_3, \mu_8, \mu_e $)
around the starting point $ \mu_3= \mu_8= 0$,
$\displaystyle\mu_e=\frac{m_s^2}{4\mu} $. \bea \Omega(\mu_3,
\mu_8,\mu_e)&\simeq &\Omega
\left(0,0,\frac{m_s^2}{4\mu}\right)+\sum_i\frac{\partial\Omega
}{\partial
\mu_i}\left(0,0,\frac{m_s^2}{4\mu}\right)\times\delta\mu_i+\frac{1}{2!}
\sum_i\frac{\partial^2\Omega }{\partial
\mu_i^2}\left(0,0,\frac{m_s^2}{4\mu}\right)\times\delta\mu_i^2\cr&+&\sum_{j\neq
l}\frac{\partial^2\Omega }{\partial \mu_j \partial
\mu_l}\left(0,0,\frac{m_s^2}{4\mu}\right)\times\delta\mu_j\delta\mu_l\label{omega
expansion} \eea where ${\mu_i}=(\mu_e,\mu_3,\mu_8)$ and
$\displaystyle\delta{\mu_i}=\left(\mu_e-\frac{m_s^2}{4\mu},\mu_3,\mu_8\right)$.
Moreover \bea
 \Omega_0 &=& -\frac{1}{12\pi^2}\sum_{\alpha=r,g,b}
 \left(\mu_{u_\alpha}^4+
 \mu_{d_\alpha}^4+2\int_0^{p^F_{s_\alpha}}
 \frac{d^3p}{(2\pi)^3}\left[\sqrt{p^2+m_s^2}-\mu_{s_\alpha}\right]\right) -
\frac{\mu_e^4}{12\pi^2}    \label{eq:OmegaNorm11222} \eea is the
free energy of the normal phase and \be
p^F_{s_\alpha}=\sqrt{\mu_{s_\alpha}^2-m^2_s}\ .\ee Besides the
inclusion of color chemical potentials, other effects that might
arise from ${\cal O}\left(1 /\mu\right)$ terms are as follows. First
we have higher order terms in the expansion of the strange quark
energy, which produces the following change of the strange Fermi
momentum, with respect to the massless case:\be
{p}^F_{s_\alpha}\simeq
\mu_{s_\alpha}-\frac{m_{s}^2}{2\mu_{s_\alpha}}-\frac{1}{2\mu}
\left(\frac{m_{s}^2}{2\mu}\right)^2 .\ee
  An effect of the next-to-leading order term in the $1/\mu$
  expansion is to change the result $\displaystyle \mu_e=\frac{m_s^2}{4\mu}$,
   which holds in the normal phase in the leading approximation.
  It is worth mentioning here that
these corrections do not affect the relation between $q_{I}$ and
$\delta\mu_{I}$: $q_{I}=1.1997\delta\mu_{I}$, because this relation
is corrected only by terms of the order  $\Delta^6$ that are
negligible within our approximation.

There are other higher order effects that can be neglected, due to
the assumption of weak coupling. First of all we can neglect modes
outside a shell around the Fermi surfaces, and therefore we assume
an ultraviolet  cutoff $\delta\ll\mu$. Furthermore we can neglect
the contribution of antiquarks, because their effect would produce a
correction of the order of $g\mu\delta$ where $g$ is the
four-fermion Nambu-Jona Lasinio coupling (of dimension $[M]^{-2}$).
In the weak coupling regime $g\mu\delta\ll g\mu^2\ll 1$ and
therefore the above mentioned approximation is justified.

\section{Results\label{sec4} }
The neutrality condition \be \frac{\partial\Omega}{\partial\mu_e}~=~
\frac{\partial\Omega}{\partial\mu_3}~=~\frac{\partial\Omega}{\partial\mu_8}~=~0
\label {eq:electrNeutra111} \ee with $\Omega$ expressed by
(\ref{omega expansion}) determines $\mu_e$ , $\mu_3$ , $\mu_8$, for
which we get the following results. For $\mu_3$ we get \be
\label{mutre}\mu_3=\,\frac {\Delta_2^2\Delta_3^2A_+}{3}\,+\,{\cal
O}\left(\frac{\Delta^4}\mu\right)+{\cal O}\left(\Delta^6\right)\
,\ee with \be A_\pm=\frac{4}{y(y^2-16q^2)}
\pm\frac{1}{2qy^2}\ln\Big|\frac{4q-y}{4q+y}\Big|\label{27}\ \ee
where\be y = \frac{m_s^2}{2\mu}\ .\ee

 We have not included a term proportional to
$\Delta_2^2\Delta_3^2\delta\mu_e$ because, as it will be clear
below, $\delta\mu_e$ contains terms of the order of at least $1/\mu$
or  $\Delta^2$. For $\mu_8$ we get \be \label{muotto}\mu_8=-\,\frac
{\Delta_2^2\Delta_3^2\,A_-}{2}\,+\,{\cal
O}\left(\frac{\Delta^4}\mu\right)+{\cal O}\left(\Delta^6\right)\
.\ee Also in this case we do not include terms proportional to
$\Delta_2^2\Delta_3^2\delta\mu_e$. Finally $\delta\mu_e$ is given by
\be \label{mue}\delta\mu_e=\frac 1
2\left[-\frac{y^2}{6\mu}+B(\Delta_2^2-\Delta_3^2)\,+\,C\,\frac{\Delta^2_2}\mu\,+D(
\Delta_2^4-\Delta_3^4)+E\Delta_2^2\Delta_3^2\right]\,+\,{\cal
O}\left(\frac{\Delta^4}\mu\right)\,+\,{\cal O}(\Delta^6) \ee where
\bea B&=&-y^2E=\frac 1 q\ln\Big|\frac{4q+y}{4q-y}\Big|\,, \cr
C&=&\frac{16y^2}{16q^2-y^2}\,,\cr D&=&\frac{64y}{(16q^2-y^2)^2}\
.\label{30}\eea In Eqs. \eqref{27} and \eqref{30} $q$ is defined by
$q=1.1997\,y/4$.
 On the basis of
the results in Eqs. \eqref{mutre}, \eqref{muotto} we conclude that
the corrections due to finite chemical potentials $\mu_3$ and
$\mu_8$ correspond to terms proportional to $\Delta^6$ or smaller in
the free energy. Since in the G-L expansion we take terms up to
order ${\cal O}(\Delta^4)$, we conclude that the chemical potentials
can be put equal to zero.

To the equations implementing color and electric neutrality we
have to add the gap equations $\displaystyle
\frac{\partial\Omega}{\partial\Delta_2}=\frac{\partial\Omega}{\partial\Delta_3}=0$.
In this way we can find the values of
$\Delta_2\,\Delta_3\,,\delta\mu_2\,,\delta\mu_3$ as functions of
$m_s^2/\mu$ for fixed values of $\mu$ ( $\mu=500$ MeV) and the CFL
gap $\Delta_0$ (25 MeV in our numerical evaluation).

 The
effect of the $\displaystyle{\cal O}\left(1/\mu\right)$ corrections
can be grasped looking at their impact on the distances between the
different quark chemical potentials. Since $\mu_3$ and $\mu_8$ are
ineffective we consider
\bea\delta\mu_{2}&=&\frac{\mu_{u_\alpha}-\mu_{s_\alpha}}2~=~\delta\mu_{u_r,s_b}=\delta\mu_{u_b,s_r}=
\frac1 2\left[\frac{m_s^2}{2\mu}+\frac{1}{3\mu}\left
(\frac{m_s^2}{2\mu}\right)^2\,-\,\mu_{e}\right]\label{deltamu2}\cr
\delta\mu_{3}&=&\frac{\mu_{d_\alpha}-\mu_{u_\alpha}}2~=~
\delta\mu_{u_r,d_g}=\delta\mu_{u_g,d_r}=\frac{\mu_{e}}{2}
\label{deltamu3}\ . \eea

\begin{figure}[ht]
\includegraphics[width=10cm]{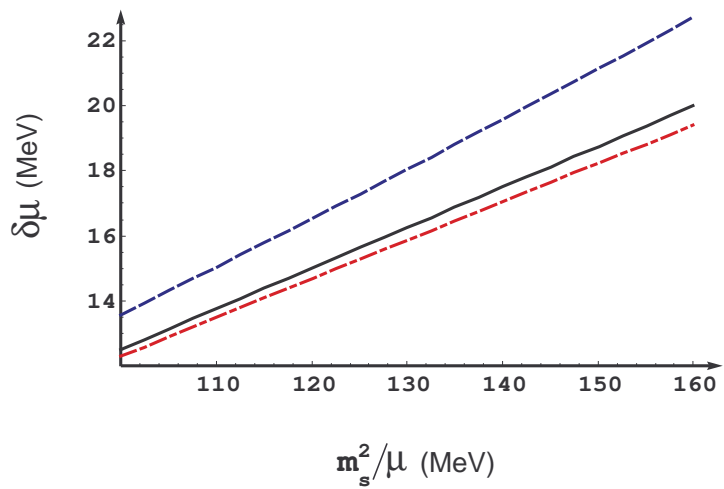}
\caption{{\rm Quark chemical potential differences $\delta\mu_{2}$
(dashed, blue online) and $\delta\mu_{3}$ (dash-dotted, red online)
together with their common value $\displaystyle\frac{m_{s}^2}{8\mu}$
obtained neglecting $1/\mu$ corrections (full line).
 \label{fig1}}}\end{figure} We note that the
effect of introducing the $1/\mu$ correction is to induce an
asymmetric splitting of the Fermi surfaces of the $s$ and $d$ with
respect to the $u$ quark Fermi sphere, more exactly
$\delta\mu_2>\delta\mu_3$; neglecting higher order effects, as in
\cite{Casalbuoni:2005zp}, one would get the result
$\delta\mu_2=\delta\mu_3$ since in this limit $\displaystyle
\mu_e=\frac{m_s^2}{4\mu}$. The results for the splitting of chemical
potentials are reported in Fig. \ref{fig1} (dash-dotted and dashed,
respectively red and blue online) together with the result obtained
in the large $\mu$ limit (continuous black line).

\begin{figure}[ht]
\includegraphics[width=10cm]{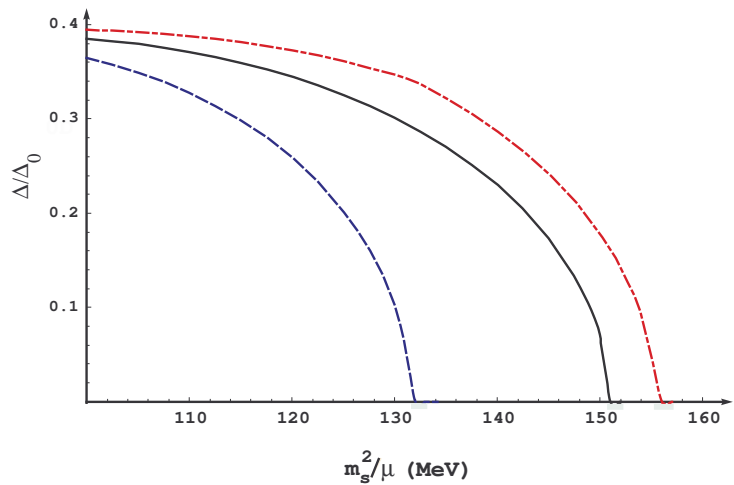}\caption{{\rm  The full
line represents the previous solution $\Delta_{2}=\Delta_{3}$; the
dashed (blue online) and dash-dotted (red online) lines are
respectively $\Delta_{2}$ and $\Delta_{3}$ in the new approximation.
All the gap parameters are normalized to the CFL homogeneous gap
$\Delta_{0}=25MeV$.
 \label{fig2}}}\end{figure}

As a consequence of these results we expect a difference between the
two gap parameters, with $\Delta_2<\Delta_3$. This is confirmed by
the results of the gap equations that are reported in Fig.
\ref{fig2}. Also in this case we report the two gaps in the new
approximation (dash-dotted and dashed, respectively red and blue
online)
 together with the result $\Delta_2=\Delta_3$ obtained
in the large $\mu$ limit (continuous black line).

 \begin{figure}[ht]
\includegraphics[width=10cm]{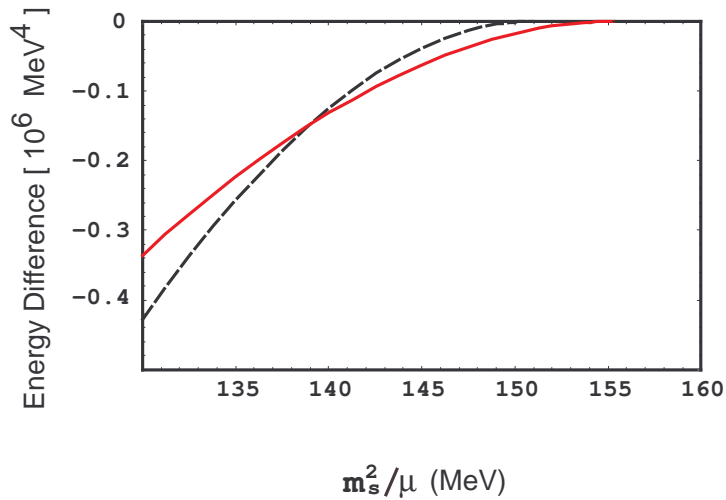}\caption{{\rm
  The free energy difference in the
previous and in the present approximation (respectively dashed black
line
 and continuous line, red online).
 \label{fig3}}}\end{figure}

From these results we can compute the difference between the free
energies in the LOFF state and in the normal phase. This result is
reported in Fig. \ref{fig3} (continuous curve, red online), together
with the result of \cite{Casalbuoni:2005zp} obtained neglecting
$1/\mu$ corrections (dotted line).

The effect of the corrections considered in the present paper is to
change the nature of the phase transitions in the LOFF regime. The
LOFF has been enlarged, as it can be seen from both figs. \ref{fig2}
and \ref{fig3}. However, whereas in the previous approximation the
LOFF window was characterized by one gap, and by condensation in two
channels: $u-s$ and $u-d$, now we have LOFF pairing in both channels
only in a tiny region, for smaller valuers of $m_s^2/\mu$, while
near the end point of the superconductive region ($m_s^2/\mu> 132$
MeV) only $\Delta_3\neq 0$ and therefore there is no $u-s$ pairing.
We can call this new phase LOFF2s. It differs from the the LOFF
phase with two flavors $u$ and $d$ for the presence of the strange
quark, which plays an active role in modifying, by mass effects, the
chemical potentials and being decisive to implement electric
neutrality. The transition between the two LOFF phases is second
order.

We finally note that we are not able to fix the lower end of the
LOFF phase with two gaps $\Delta_2\neq 0$ and $\Delta_3\neq 0$. As a
matter of fact this point would be obtained by comparing the LOFF
and the gCFL free energies, which will be only feasible going beyond
the existing approximation, when all the $1/\mu$ corrections  in the
gCFL phase are considered. In any case we expect that the LOFF phase
with two gaps, in the one plane wave approximation, is limited to a
narrow range of values of $m_s^2/\mu$, because, in the previous
approximation, its lower end was for $m_s^2/\mu\sim 128$ MeV
\cite{Casalbuoni:2005zp} and LOFF2s begins at $m_s^2/\mu\sim 132$
MeV .

\section{Conclusions\label{sec5} } An interesting point to be investigated is
the chromomagnetic stability of the LOFF2s phase. We have applied
the results of \cite{Ciminale:2006sm} to this phase, characterized,
as we have stressed by $\Delta_2=0$ and $\Delta_3\neq 0$ and we have
found that it is chromomagnetically stable: three gluons, related to
the generators of an unbroken $SU(2)_c$  remain massless while all
the Meissner masses of the remaining gluons (both the longitudinal
and transverse ones) are real, a result that is also implicit in the
conclusions of refs. \cite{Giannakis:2005vw}. Incidentally, since
there is change of symmetry, this result shows that LOFF2s and the
LOFF state with two nonvanishing gaps are distinct phases. It would
be tempting to extend the analysis of stability  to the tiny region
with $m_s^2/\mu$ near, but smaller of $132$ MeV, and $\Delta_2\neq
0$, $\Delta_3\neq 0$. This extension is complicated by the fact that
we do not know exactly where this LOFF phase extends, because a
complete ${\cal O}(1/\mu)$ calculation of the gCFL has not yet been
performed. We plan to come back to this problem in the future.

In conclusion we have proved the relevance of $1/\mu$ effects for
the LOFF phase with three flavors in the one-plane wave
approximation. Some of our results can be important also for the
more complex studies with several plane waves
\cite{Rajagopal:2006ig,Rajagopal:2006dp}. Since also in this case
the inclusion of the $1/\mu$ contributions breaks the symmetry
between $d-u$ and $u-s$ pairings, this can provide some hints on the
best strategy to follow in order to identify the most favored
crystalline structures. Moreover we have found that at the order
$\Delta^6$ the color chemical potentials do not vanish, which can be
also important in crystallography of three-flavor quark matter
because in the GL expansion one has to take into account terms up to
this order.

\end{document}